\documentstyle [psfig,12pt] {article}

\newcommand{\sect}[1]{ \section{#1} \setcounter{equation}{0} }
\newcommand{\req}[1]{(\ref{#1})}
\newcommand{\nwc}{\newcommand}

\newcommand{\wls}{Wilson lines}

\nwc{\gev} {{\rm GeV}}
\nwc{\tev} {{\rm TeV}}
\nwc{\mP} {$M_{\rm Planck}$}
\nwc{\mx} {$M_{\rm X}$}
\nwc{\ms} {$M_{\rm string}$}
\nwc{\sieb}{\bf \overline{27}}

\newcommand{\Z}{\ZZ}
\def\bfone{\relax{\rm 1\kern-.35em 1}}
\def\inbar{\vrule height1.5ex width.4pt depth0pt}
\def\IC{\relax\,\hbox{$\inbar\kern-.3em{\mss C}$}}
\def\ID{\relax{\rm I\kern-.18em D}}
\def\IF{\relax{\rm I\kern-.18em F}}
\def\IH{\relax{\rm I\kern-.18em H}}
\def\II{\relax{\rm I\kern-.17em I}}
\def\IN{\relax{\rm I\kern-.18em N}}
\def\IP{\relax{\rm I\kern-.18em P}}
\def\IQ{\relax\,\hbox{$\inbar\kern-.3em{\rm Q}$}}
\def\IR{\relax{\rm I\kern-.18em R}}
\def\ZZ{\relax{\hbox{\mss Z\kern-.42em Z}}}
\font\cmss=cmss10 \font\cmsss=cmss10 at 7pt
\def\ZZ{\relax\ifmmode\mathchoice
{\hbox{\cmss Z\kern-.4em Z}}{\hbox{\cmss Z\kern-.4em Z}}
{\lower.9pt\hbox{\cmsss Z\kern-.4em Z}}
{\lower1.2pt\hbox{\cmsss Z\kern-.4em Z}}\else{\cmss Z\kern-.4em
Z}\fi}
\nwc{\hyp} {\hyphenation}
\nwc{\be}  {\begin{equation}}
\nwc{\ee}  {\end{equation}}
\nwc{\ba}  {\begin{array}}
\nwc{\ea}  {\end{array}}
\nwc{\bdm} {\begin{displaymath}}
\nwc{\edm} {\end{displaymath}}
\nwc{\pr}  {\prime}
\nwc{\bea} {\be\ba{lcl}}
\nwc{\eea} {\ea\ee}

\nwc{\bda} {\bdm\ba{lcl}}
\nwc{\eda} {\ea\edm}

\nwc{\bc}  {\begin{center}}
\nwc{\ec}  {\end{center}}

\nwc{\ds}  {\displaystyle}
\nwc{\bmat}{\left(\ba}
\nwc{\emat}{\ea\right)}
\nwc{\nn} {\nonumber}
\nwc{\nnn} {\nonumber \vspace{.2cm} \\ }
\nwc{\ra}{\rightarrow}
\nwc{\lra}{\longrightarrow}

\nwc{\p} {\partial}

\nwc{\scr}  {\scriptstyle}
\nwc{\tx}  {\textstyle}
\nwc{\scs} {\scriptscriptstyle}

\nwc{\ov}  {\overline}

\nwc{\hb}  {\bar h}
\nwc{\xb}  {\bar x}
\nwc{\yb}  {\bar y}
\nwc{\zb}  {\bar z}
\nwc{\wb}  {\bar w}
\nwc{\Ob}  {\bar O}
\nwc{\Yb}  {\bar Y}

\nwc{\ep} {\epsilon}
\nwc{\de} {\delta}
\nwc{\Th} {\Theta}
\nwc{\th} {\theta}
\nwc{\al} {\alpha}
\nwc{\si} {\sigma}
\nwc{\Si} {\Sigma}
\nwc{\om} {\omega}
\nwc{\Om} {\Omega}
\nwc{\Ga} {\Gamma}
\nwc{\ga} {\gamma}
\nwc{\bet} {\beta}
\nwc{\La} {\Lambda}
\nwc{\la} {\lambda}

\nwc{\Sc}  {{\cal S}}
\nwc{\Rc}  {{\cal R}}
\nwc{\Dc}  {{\cal D}}
\nwc{\Oc}  {{\cal O}}
\nwc{\Cc}  {{\cal C}}
\nwc{\gc}  {{\cal g}}

\nwc{\Of}  {{\cal O}_f}
\nwc{\Oft} {{\cal O}_{f_2}}
\nwc{\Ofo} {{\cal O}_{f_1}}

\nwc{\Pc}  {{\cal P}}
\nwc{\Mc}  {{\cal M}}
\nwc{\Ec}  {{\cal E}}
\nwc{\Fc}  {{\cal F}}

\nwc{\Hc}  {{\cal H}}
\nwc{\Kc}  {{\cal K}}
\nwc{\Wc}  {{\cal W}}

\nwc{\Fcp} {{\cal F}^\pr}
\nwc{\Hcp} {{\cal H}^\pr}

\nwc{\Xc}  {{\cal X}}
\nwc{\Gc}  {{\cal G}}
\nwc{\Zc}  {{\cal Z}}
\nwc{\Nc}  {{\cal N}}

\nwc{\xc}  {{\cal x}}
\nwc{\Ac}  {{\cal A}}
\nwc{\Bc}  {{\cal B}}
\nwc{\Uc} {{\cal U}}
\nwc{\Vc} {{\cal V}}
\nwc{\Lc} {{\cal L}}
\nwc{\Qc} {{\cal Q}}

\nwc{\lng} {\langle}
\nwc{\rng} {\rangle}

\nwc{\lf} {\left}
\nwc{\ri} {\right}

\nwc{\diag} {{\rm diag}}
\nwc{\inv}  {{\rm inv}}
\nwc{\mod}  {{\ \rm mod\ }}
\nwc{\dete}  {{\rm det}}
\nwc{\tr}  {{\rm tr}}
\nwc{\im}  {{\rm Im}}
\nwc{\re}  {{\rm Re}}

\nwc{\h} {\frac{1}{2}}
\nwc{\fc} {\frac}

\def\KK{\relax{\rm I\kern-.18em K}}
\def\RR{\relax{\rm I\kern-.18em R}}
\def\NN{\relax{\rm I\kern-.18em N}}
\def\PP{\relax{\rm I\kern-.18em P}}

\def\zz{\relax{\sf Z\kern-.3em Z}}
\def\ZZ{\relax{\sf Z\kern-.4em Z}}
\def\ZZZ{\relax{\sf Z\kern -.5em Z}}
\def\ZZZ{Z\kern -0.37em Z}
\def\QQ{{\rm \kern .25em
             \vrule height1.4ex depth-.12ex width.06em\kern-.31em Q}}
\def\CC{{\rm \kern .25em
             \vrule height1.4ex depth-.12ex width.06em\kern-.31em C}}
\if@twoside\oddsidemargin25mm
\else\oddsidemargin25mm\evensidemargin25mm\marginparwidth25mm\fi%
\begin{document}
\begin{titlepage}

\vskip 6cm

\title{$\ $\ \\$\ $\ \\ $\ $\ \\$\ $\ \\Moduli Dependence of One--Loop Gauge
Couplings in (0,2) Compactifications\thanks{Supported by the
 "Sonderforschungsbereich 375--95 f\"{ur} Astro--Teilchenphysik" of the
Deutsche Forschungsgemeinschaft and the EEC under contract no.
SC1--CT92--0789.}}

\vskip 3cm

\author{{P. Mayr$^1$\ } {\small and} {\ S. Stieberger$^2$}\\ \\ \\
$^1${\em Theory Division, CERN}\\
{ \em CH--1211 Geneva 23, SWITZERLAND}\\[1cm]
$^2${\em Institut f\"{u}r Theoretische Physik} \\
{\em Physik Department} \\
{\em Technische Universit\"at M\"unchen} \\
{\em D--85747 Garching, FRG}}

\date{}
\maketitle
\thispagestyle{empty}
\vskip 2cm
\begin{abstract}
We derive the moduli dependence of the one--loop gauge couplings
for non--vanishing gauge background fields in a four--dimensional
heterotic (0,2) string compactification. Remarkably, these functions turn
out to have a representation as modular functions on an auxiliary
Riemann surface on appropriate truncations of the full moduli
space.
In particular, a certain kind of one--loop functions is given by the
free energy of two--dimensional solitons on  this surface.
\end{abstract}

\begin{picture}(2,2.5)(-360,-630)
\put(12,-95){CERN--TH/95-96}
\put(12,-110){TUM--HEP--208/95}
\put(12,-125){hep-th/9504129}
\put(12,-145){April 1995}
\end{picture}

\end{titlepage}

\sect{Introduction}
One loop gauge couplings in string theories have been a subject of
alive interest over the last four years for two obvious reasons
\cite{kap1,dkl2,thr1}.
Firstly, these so--called threshold functions
represent the boundary conditions for the running
gauge couplings of the effective field theory at the string scale
and determine in this way the values of the low--energy gauge couplings
\cite{kap1}.
Therefore their knowledge is a basic ingredient for string phenomenology.
Secondly, their dependence on the vacuum expectation values
of the moduli fields is of great theoretical interest: it defines
the largest possible subgroup of the tree--level target space duality
symmetries \cite{sti1} which is realized in the quantum theory.
It is important to keep in mind this relation: in general,
the thresholds can restrict the quantum
symmetries and cannot be determined by imposing the tree--level
symmetries \cite{ms1}.

Recently, the moduli dependence of the one--loop couplings gained in
importance in the light of the fascinating subject of S--duality
in N=2 supersymmetric field theories \cite{sd1}. The discussion
of strong--weak coupling symmetries as true symmetries of string theories
or relations between them
requires the knowledge of the moduli dependence
of the gauge couplings. In particular, this dependence
determines the classical and perturbative monodromies in
N=2 supersymmetric theories \cite{antn,dwit}. In fact as we will explain below
the moduli dependent part of the one--loop couplings in the N=1 theory
agrees with the corresponding couplings in a closely related N=2 theory.
Therefore our results can be directly applied to the above problem
for this specific N=2 theory.

Roughly speaking there are two types of moduli in a four--dimensional
string theory with an interpretation as a compactified ten--dimensional
string theory:
the moduli which describe the geometry and complex structure
of the six compactified internal dimensions and the Wilson line moduli
in the gauge sector which have the interpretation of flat
but homotopically non--trivial gauge connections wrapping around the
non--trivial cycles of the compactification manifold. While the
dependence of the one--loop gauge coupling on the first type has been
considered in great detail in the past, the same cannot be said about the
Wilson line type of moduli. The only exactly known
Wilson line dependent gauge couplings are restricted to theories, where
the background gauge fields are quantized \cite{ms3}. While the string theories
considered there are of phenomenological relevance, no information
about the global structure of the moduli space of continuously connected
string vacua can be drawn. Lowest order results in an expansion in
the vacuum expectation values of the Wilson moduli have been obtained
in \cite{antmu} by a string computation and more
recently\footnote{see also ref. \cite{dwit}.} in \cite{clm2}
using the concept of the topological free energy. It is clear that such
an expansion naturally misses the global structure of moduli space.
In addition, technical difficulties in the latter approach
make necessary an unsystematic truncation even in the lowest order
of the background values and an assumption about the duality symmetry
of the one--loop couplings rather than a proof of it.

To close this gap, we will derive the duality symmetries and functional
dependence of Wilson line dependent gauge couplings in an orbifold theory
\cite{orbi1}
with a twist embedding of the space group.
Interestingly, our results show a strong connection to 2d physics on
an auxiliary Riemann surface combining the metric and gauge moduli
in a common setting.

Another interesting implication of the presence of moduli
dependent threshold  corrections in general
is the fact that those subspaces of moduli space which enter
the one--loop couplings have to be compatible with N=2 space--time
supersymmetry \cite{msp}. The geometry of such subspaces is closely related
to the special geometry of the N=2 theory which consists only of the N=2
twisted and the untwisted sector. In fact the Wilsonian gauge couplings
and the K\"ahler potential of these special moduli fields can be obtained
from a holomorphic prepotential $F$ in the usual way. The duality
transformations become a subset of the symplectic transformations acting
on the symplectic vectors $(X^A,F_A)$ of N=2 supergravity \cite{cer1}.
The situation is similar to the one which has been used to infer the
special geometry structure of the CY moduli spaces in the case of
vanishing Wilson lines \cite{cy2}.
While in these cases the N=2 theory whose consistency
implied the special geometry of the moduli subspace
was the compactification of a type II superstring
theory, in the present case this r\^ole is taken over by the N=2 theory which
is related to the N=1 theory by omitting the N=1 sectors. This theory
appropriately includes the vector multiplets whose scalar components
are the Wilson line moduli which could not be described by a type II
compactification.

\sect{Symmetries and string amplitudes of the orbifold theory}
Our model is a
$\Z_8$ orbifold
defined on  the six--dimensional torus lattice
$\La_6=SO(4)\times SO(9)$. In the complex basis the twist has the eigenvalues
$\th=\exp[\fc{2\pi i}{8}(-4,1,3)]$. The twist embedding in the gauge lattice
$\La_{16}
=E_8 \times E_8'$ is chosen to be $\Th= \Z_2^{(1)} \times \Z_2^{(1)} \times
\Z_{8}^{(5)} \times \Z_4'^{(3)} \times \Z_2'^{(3)}$ \cite{holo}.
The resulting gauge group for zero values of the Wilson lines
is $SU(6) \times SU(2) \times U(1)^2 \times SU(4)'\times SU(3)'\times
SU(2)'^2\times U(1)'$. In addition, we introduce two complex
non--vanishing Wilson lines in the first $E_8$:
\be
a_1^I=(\la_1,\la_2;0,0;0,0;0,0)\ \ \ ;\ \ \
a_2^I=(\mu_1,\mu_2;0,0;0,0;0,0)\ \ \ {\rm with}\ \ \ \la_i,\mu_i \in \RR\ ,
\ee
where the entries are w.r.t to the two  weights $d_1=(1/\sqrt 2,0),
d_2=(0,1/\sqrt 2)$ of $SU(2)^2$.
The sublattice $SO(4) \simeq SU(2)\times SU(2)$ remains unrotated in the N=2
sector which consists of all boundary conditions along the two cycles
of the world--sheet torus created out of $1,\th^2,\th^4$ and $\th^6$.
The windings and momenta are denoted by
$n^1,n^2$ and $m_1,m_2$, respectively.
The complex moduli fields which belong to this plane and contain the \wls\
can be defined along \cite{clm1}:
\bea \label{compmod}
U&=&\ds{\fc{R_2}{R_1}e^{i \phi}}\ ,\nnn
T&=&\ds{\tilde{T}-\fc{1}{4}(\la_1\mu_1+\la_2\mu_2)+
\fc{1}{4}U(\la_1^2+\la_2^2)}\ ,\nnn
B&=&\ds{\h(\mu_2+i\mu_1)-\h U(\la_2+i\la_1)}\ ,\nnn
C&=&\ds{\h(-\mu_2+i\mu_1)-\h U(-\la_2+i\la_1)\ ,}
\eea
with $\tilde{T}=2 b+2 i R_1 R_2 \sin \phi$ being
the K\"ahler--modulus without \wls. $R_1$
and
$R_2$ are the radii of the two underlying $SU(2)$ root lattices, respectively
and $\phi$ is their relative orientation. The four--dimensional subspace
of the $E_8$ which is left
fixed under $\Th^2$ is an $SO(8)$ root lattice described by the set of
quantum numbers $k=(k_1,k_2,k_3,k_4)$ and the metric $g_{SO(8)}$
with $k^tg_{SO(8)}k=2(k_1^2-k_1k_2+k_2^2-k_2k_3-k_2k_4+k_3^2+k_4^2)$
\cite{msp}. There are various possibilities for the gauge
groups at special  values of the gauge background fields.

The starting point to calculate the non--universal part of the
one--loop threshold corrections
$\triangle_a$ to the inverse gauge coupling $g_a^2$
is the general formula
of ref. \cite{kap1}. Only the N=2 sector of an orbifold gives rise to
moduli dependent threshold corrections \cite{dkl2}. For a generic
gauge group factor the general formula can be simplified
to \cite{a,msp}:
\be
\triangle(T ,\bar{T },U,\bar U,B,\bar B,C,\bar C)=
\fc{1}{3}b_0^{N=2}\int\limits_{\tilde \Fc} \fc{d^2 \tau}{\tau_2}
Z_{SU(2)^2 \times E_8^{\rm inv.}}^{(1,\Th^2)}(\tau,\bar \tau
)\ \Cc_{(1,\Th^2)}(\tau)\ ,
\label{th8}
\ee
with: \small
\be
\ba{rcl}
\ds{Z^{(1,\Th^2)}_{SU(2)^2 \times E_8^{\rm inv.}}(\tau,\bar \tau)}&=&
\ds{\sum_{n^1,n^2 \in \ZZZ
\atop m_1,m_2 \in \ZZZ}\sum_{k \in \ZZZ^4}e^{\pi i \tau\lf[2m_1n^1+2m_2n^2+
k^tg_{SO(8)}k\ri]}e^{-2\pi \tau_2|p_R|^2}}\ ,\nnn
\ds{\Cc^{-1}_{(1,\Th^2)}(\tau)}&=&\ds{\sum_{l\in\ZZZ^4}
e^{\pi i\tau l^tg_{SO(8)}l}\ ,}\nnn
p_R&=&\textstyle \frac{1}{\sqrt{2\im \tilde{T} \im U}}
\scriptstyle \left[(T  U+BC)\
n^2+T n^1-Um_1+m_2+i(B+C)(k_1-\fc{1}{2}k_2)-\h(B-C)k_2\right]
\displaystyle \ .
\ea
\ee\normalsize
Here, $b_0^{N=2}$
is the $\beta$--function coefficient of the $N=2$ sector for the case
$B$=$C$=0. The region of integration is extended to
$\tilde \Fc=\{1,S,ST\}\Fc_1$
to take into account\footnote{This is explained in more detail
in \cite{ms1}.} the {\em different} contributions of the twisted sectors.
Note that the plane $(k_1,k_2)$
is not orthogonal to the plane $(k_3,k_4)$. On the other hand, this is
important to ensure the integrand to be invariant\footnote{Modular invariance
is important to reproduce the full moduli dependence as
well as to calculate the topological free energy.}
under $\Ga_0(2)_\tau$. The analogous expressions for gauge groups
with additional massless charged particles for special values of the
moduli fields can be found in \cite{msp}.

It can be proven that \req{th8} is invariant under the following
transformations together with unimodular transformations on the momentum and
winding numbers: Firstly, there are the generalizations of the
$SL(2,\Z)_{T }\times SL(2,\Z)_U$ transformations

\begin{eqnarray} \label{s1}
\ds{T } &\lra&\ds{-\fc{1}{T }\ ,\ B\lra \fc{B}{T }\ ,\ C\lra
\fc{C}{T }\ ,\ U \lra U+\fc{BC}{T }}\ ,\\
T &\lra&T +1\ , \\
\ds{U}&\lra&\ds{-\fc{1}{U}\ ,\ B\lra \fc{B}{U}\ ,\ C\lra
\fc{C}{U}\ ,\ T  \lra T +\fc{BC}{U}}\ ,\\
U&\lra&U+1\ ,
\end{eqnarray}
Moreover there are the two kinds of shifts acting on the Wilson lines:
\begin{eqnarray}
\label{s2}
B&\lra& B+i r+s\ ,\ C\lra C+ir-s\ ,\\[4mm]
B&\lra& B-iUx-Uy,\qquad C\lra C-iUx+Uy,\\[1mm]
T  &\lra& \ds{T
+(x^2+y^2)U-y(B-C)+ix(B+C)}\ ,\nonumber
\label{s3}
\end{eqnarray}
with $r,s,x,y \in \Z$.
In addition, there is a mirror transformation exchanging $T $ and $U$, a
symmetry under the exchange of $B$ and $C$ and a parity transformation:
\begin{eqnarray}
T &\longleftrightarrow& U\ ,\\[4mm]
B &\longleftrightarrow& C\ ,\\[4mm]
B\lra -B&,&C\lra -C\ .
\label{s31}
\end{eqnarray}
The transformations \req{s1} -- \req{s31}  represent a (non--minimal)
set of generators for the duality symmetries of the moduli
dependent gauge couplings given in \req{th8}.
\ \\
\sect{Automorphic functions}
Having determined the symmetries of the one--loop gauge couplings
depending on the four--moduli subspace parametrized by $T,\ U,\ B$ and
$C$ we will derive now their functional dependence. In addition to the
symmetries we have to impose boundary conditions specifying the
behavior of the functions
at the locus of the moduli space ${\cal M}$ where additional particles
become massless, causing singularities of the effective field theory.
In fact, we will reduce the moduli space once more by restricting to
the subspace  ${\cal M}_i:\
\lambda_i,\mu_i=0$ for $i=1\ \vee \ i=2$ and consider for the moment
this three moduli
problem. The K\"ahler potential \cite{wit1,fer1,clm1} can be written
as $K=-\ln Y$ with
\be \label{kp001}
Y=(T-\bar{T})(U-\bar{U})+(B+\bar{C})(\bar{B}+C)\ =\ \det(M-M^\dagger) \ ,
\ee
where
\be \label{kp002}
M  = \left( \ba{ll} T & B\\ -C & U \ea \right) \ .
\ee
A $Sp(4,\Z)$ subgroup of the duality symmetries \req{s1} -- \req{s31}
is realized on $M$ by the action
\be \label{sp4_1}
M \rightarrow (A\ M + B)\ (C\ M + D)^{-1} \ , \qquad
\left( \ba{ll} A & B\\C & D \ea \right) \in Sp\ (4,\Z) \ .
\ee
Indeed on ${\cal M}_i$ we can define the new moduli $B_i =
\fc{1}{2} \mu_i - \fc{1}{2} U \lambda_i$ and the transformations
\req{sp4_1} acting on the matrices
\be \label{mis}
M_i = \left( \ba{ll} T & B_i\\ B_i & U \ea \right)
\ee
becomes identical to
the standard action of $Sp(4,\Z)$ on an element of the Siegel upper half
plane $\Sc _2$. Note that the defining condition $\im\ M_i > 0$ is ensured
by the positivity of both $\im\ T$ and $-Y$.

In fact, the above treatment is motivated by a non--trivial property of
the moduli dependent one--loop gauge couplings $\triangle_a$ for vanishing
Wilson lines obtained in \cite{dkl2}:
\be \label{oldth}
\triangle_a \sim \ln (T-\bar{T})|\eta_T|^4\ (U-\bar{U})|\eta_U|^4 \ ,
\ee
where $\eta_T=\eta(T),\ \eta_U$ are Dedekind's functions.
The non--holomorphic piece proportional to the K\"ahler potential
is of pure field--theoretical origin and arises from the coupling of
the fermions to the K\"ahler and sigma--model connections \cite{lou1,der1}.
Eq. \req{oldth} is nothing but the sum of free energies
of solitonic configurations \cite{alv1}
of a complex boson on a
Riemann surface given by the product of two tori with Teichm\"uller
parameters $T$ and $U$, respectively (fig. 1a):
\be
\triangle_a \sim \sum_\alpha \ \ln\  \left[(\im\ T)^\h |\theta_\alpha(T)|^2
\right]
\ + \ \sum_\alpha \ \ln\ \left[(\im\ U)^\h |\theta_\alpha(U)|^2\right]\ .
\ee
Here the $\theta$'s are the even $g=1$ Riemann theta constants
and the sum is over the CP even boundary conditions
along the two plus two cycles of the
product of two tori. Amazingly, this interpretation has an immediate
generalization to the above defined three moduli problem with non--vanishing
Wilson line. Combining $T,\ U$ and $B_i$ as in \req{mis} we consider the
sum over the free energies
of a complex boson on a genus two Riemann surface $K$ with
period matrix $M_i$ (fig.1b):
\be \label{sol2}
\sum_\alpha \ \ln \left[(\det \im\  M_i)^\h |
\vartheta_\alpha(M_i)|^2 \right] \ ,
\ee
where now the $\vartheta$'s are the ten even $g=2$ theta constants and
the summation is over the CP even boundary conditions along the nontrivial
cycles of $K$.
As we will argue this is precisely the functional dependence of the
threshold function w.r.t. a gauge group coupled to charged particles
which become massless for $B_i=0$ corresponding to the degeneration
limit shown in fig.1c ! Moreover
note that the direct product of tori is resolved into a common Riemann
surface, a circumstance
which is certainly necessary for a possible two--dimensional
interpretation of the other kinds of threshold functions considered below.

We will make use now of the theory of Siegel modular forms of degree
two to determine the functional dependence of the gauge coupling functions
much in the same way as it can be done in the one--moduli case
with the modular forms of degree one playing the central r\^ole.
To clarify the philosophy let us sketch briefly
the ingredients from which $SL(2,\Z)$ invariant threshold functions depending
on one modulus $T$ can be derived. The relevant facts are: i) the graded ring
of modular forms of $SL(2,\Z)$ is generated by two modular forms $\Ec_4$ and
$\Ec_6$ of weight 4 and 6, respectively\footnote{See e.g. \cite{kob1}.}
ii) there is a unique cusp form
$\Cc_{12}$ of weight 12 without zeros or poles  inside the Siegel
fundamental
region $\Fc_1$ and iii) any modular invariant function can be written
as a rational function $F(j)$
in the $j$--invariant. In fact, one can argue that this information is
sufficient to determine a $SL(2,\Z)$ invariant function of a given
divisor uniquely up to a constant.
The various physical boundary conditions can be distinguished corresponding
to the presence or absence of holomorphic anomalies and exceptional
massless charged states at some special points in $\cal M$, respectively.
Holomorphic anomalies lead to a non--holomorphic piece in the coupling function
as in \req{oldth} and therefore the holomorphic part will have to transform
covariantly rather than to be invariant under the duality transformations.
Exceptional massless states require zeros (or poles) inside the fundamental
domain leading to the expected logarithmic singularity in the gauge coupling.
In addition, there is the singularity in the decompactification limit
$T \ra i\infty$ (and its mirror partner $U \ra i\infty$) where
the gauge couplings can diverge as
$\lim_{T\rightarrow i \infty} \triangle_a \sim T-\bar{T}$
due to the infinite number of light Kaluza Klein states.

Without giving the mathematical details \cite{msp} let us describe the relevant
solutions for the three moduli case corresponding to $g=2$.
The ring of modular forms of $Sp(4,\Z)$ is
generated by two modular forms $\Ec_4$ and $\Ec_6$ and three cusp
forms $\Cc_{10},\ \Cc_{12}$ and $\Cc_{35}$, where again the subscript denotes
the modular weight \cite{igu1}.
In the following we will omit the odd generator
$\Cc_{35}$ since it factorizes into expressions of the same kind due to the
algebraic dependence of its square on the generators of the modular forms
of even weight. The description of these modular forms
in terms of genus two
theta--functions has been given in \cite{igu1}.
The physical boundary conditions we consider are the decompactification
limits $\lim_{T\rightarrow i \infty},\
\lim_{U\rightarrow i \infty}$,
the behavior for $B_i\ra0$,
the covariant or invariant transformation properties w.r.t. $Sp(4,\Z)$
and exceptional massless states inside $\Fc_2$.

\noindent 1. {\em Thresholds with a singularity only for $B_i
\rightarrow 0$}:
In this case there are particles charged under the gauge group under
consideration which become massless for vanishing Wilson lines.
On the other hand, there are no other singularities in the
Siegel fundamental domain $\Fc_2$ apart from the decompactification
singularities. The modular form with these
properties is $\Cc_{10}$ and the thresholds are given by the expression
\be \label{del1}
\triangle^{I}_a = \fc{b_a^{I}}{10} \ln Y^{10} |\Cc_{10}|^2 =
\fc{b_a^{I}}{10} \ln Y^{10} \lf|\prod_{k=1}^{10} \vartheta_k(0,M_i)\ri|^4 \ ,
\ee
where $b_a^{I}$ is a model--dependent
beta--function coefficient, $Y$ is given in \req{kp001}
and the $\vartheta_k$ are the ten even theta--functions at genus two.
In the limit $B_i\ra  0$ the expansion of $\triangle_a^{I}$ becomes
up to order ${\cal O}(B_i^4)$ terms:
\be
\triangle^{I}_a \rightarrow b_a^{I} \left( \ln\ Y + \ln \ |\eta_T\ \eta_U|^4
+ \frac{1}{5} \ln \ |1+6 B_i^2\ \partial_T \ln \eta^2_T \ \partial_U \ln
\eta^2_U|^2
\ | \eta_T\ \eta_U |^4 \ |B_i|^2 \right)\ .
\label{del1l}
\ee
Eq. \req{del1} is just \req{sol2} times a constant.

\noindent 2. {\em Thresholds without singularities in $\Fc_2$}:
This is the generic case where there are no points inside $\Fc_2$ with
additional charged particles w.r.t. the considered gauge group.
The unique cusp form with this behavior is $\Cc_{12}$ and the
thresholds read:
\be \label{del2}
\triangle_a^{II} = \fc{b^{II}_a}{12}\ln Y^{12} |\Cc_{12}|^2 \ .
\ee
In the limit $B_i \ra  0 $ we find the following expansion:
\be \label{del2l}
\triangle_a^{II} \ra   \fc{b^{II}_a}{12}\ln Y^{12}|\eta_T^{24}\eta_U^{24}|^2
|1+12 B_i^2 \partial_T \ln \eta^2_T \ \partial_U \ln \eta^2_U|^2 \ ,
\ee
in agreement with the result of \cite{antmu}.

\noindent 3. {\em Thresholds with special singularities in $\Fc_2$}:
In this case there are additional massless states at finite points in $\Fc_2$.
For $B_i \neq 0$ these singularities can move in the moduli space
but still exist \cite{clm2}. Since the exact expression depends on the
spectrum let us consider the simplest example of the $\Z_2$ twisted plane
in our $\Z_{8}$ model. For $B_i$=0 there are enhanced gauge symmetries for the
points in moduli space where $T=U$, $T=U=i$,$T=U=e^{2\pi i/3}$ with
extra gauge group factors $U(1),\ U(1)^2$ and $SU(2)$, respectively
\cite{ver1}. A threshold function describing
the extra massless states at these points has been proposed
in \cite{clm2} based on symmetry assumptions:
\be \label{del31}
\triangle_a \sim  \ln (j_T-j_U) = \ln\left(
\left[\h\sum_\alpha \left(\frac{\theta_\alpha(T)}{\eta(T)}\right)^8\right]^3-
\left[\h\sum_\alpha \left(\frac{\theta_\alpha(U)}{\eta(U)}\right)^8
\right]^3\right) \ ,
\ee
which has indeed the correct number of zeros at
the special points to describe the massless spectrum\footnote{Rigorous
proofs of \req{del31} have been given recently in \cite{antn,dwit}.}.
The following function combines the $B_i=0$ limit \req{del31} with
$Sp(4,\Z)$ invariance:
\be \label{del32}
\triangle_a^{III}  = b_a^{III} \ln  \fc
{\Cc_{12}^2-4\Ec_4^3\Ec_6^2+243\ \Cc_{10}\Ec_4^2 \Ec_6}
{(\Ec_4^3+\Ec_6^2-\Cc_{12})^2} \ .
\ee
The $B_i\ra  0$ expansion is
\be
\triangle_a^{III}\ra  b_a^{III} \ln \left[(j_T-j_U)^2 + 2 B_i^2
(j_T-j_U)(
\partial_U \ln \eta^2_U\ j^\pr_T -
\partial_T \ln \eta^2_T\ j^\pr_U)\right] \ ,
\label{del3l}
\ee
where $j_T^\pr=\partial_T\ j_T$ etc. Although the modular form \req{del32}
is a good candidate for the Wilson line dependent one--loop
coupling for the enhanced gauge symmetries from the six--dimensional
compactification, we stress that in this case we have no proof at the moment
that it has the correct singularity structure at higher orders in an
expansion in $B_i$. The reason is that due to the complicated singularity
structure of $\triangle^{III}_a$ its factorization as in \req{del32}
could involve modular forms of higher weight introducing a small
additional number of available coefficients. We hope to clarify the
situation in \cite{msp}.

Consider now the functional dependence of \req{del31} and its generalization
\req{del32}. From a two--dimensional
point of view \req{del31} hardly has a natural
interpretation since it correlates $T$ and $U$ dependent quantities
defined on the direct product manifold of fig.1a. However we have seen
that the correct picture in the presence of deformations in the direction
of the marginal perturbations $B_i$ is that of the genus 2 Riemann surface
in fig.1b. with the limit $B=0$ corresponding to the degeneration of
pulling the two tori away from each other (fig.1c). It is tempting to interpret
the massless states at the points
$T=U$, $T=U=\sqrt{-1-BC}, T=U=-\h+\sqrt{-\fc{3}{4}-BC}$
in terms of a 2d theory on an
exceptionally symmetric surface $K$. In terms of $K$ the above relations
on the moduli reflect symmetries of the period matrix and imply the
vanishing of the integrals of abelian differentials along certain cycles
of $K$. We have indicated the relation of the $j$--function to the partition
function of $E_8$ solitons in \req{del31}.

We have seen that we can describe the functional dependence of the
Wilson line dependent thresholds in a codimension 1 subspace of the
original moduli space (which already was a reduction of the full
moduli space of the $\Z_2$ plane) by the modular forms of genus two.
Interestingly, there is another
subspace of the moduli space which can be described by
elliptic functions of genus one. First observe that in the degeneration
limes $T \ra  i \infty$ the threshold corrections of the first kind become
\be
\triangle^{I}_a \rightarrow b_a^{I} \left(
\ln \ Y + \ln\ |\eta_U|^4 + \frac{1}{5} \ln \
\left| \frac{\theta_1(B_i,U)}{\eta_U} \right| ^2 + \frac{i\pi}{5}
(T-\bar{T})\right) + {\cal O}(e^{-2\pi T_2}) \ ,
\ee
where $T_2={\rm Im}T$  and $\theta_1$ is the single odd theta--function at
genus one.
There exists also the analogous limit for $U \rightarrow i \infty$.
Interestingly, we can write for these limits an invariant expression for
$B \neq \pm C$:
\be
\fc{\triangle^{I^\prime}_a}{b_a^{I^\prime}}=
\ln Y + \ln |\eta_U|^4
+ \frac{1}{4} \left(
\ln\left|\frac{\theta_1(\frac{B-C}{2},U)}{\eta_U} \right| ^2 +
\ln\left|\frac{\theta_1(\frac{i(B+C)}{2},U)}{\eta_U} \right| ^2 +
i \pi (T-\bar{T}) \right) + {\cal O}(e^{-2\pi T_2}) \ .
\label{tinfty}
\ee
This expression is determined uniquely by the singularities and the
global symmetries.
In fact, an expansion in powers of $q_T = e^{2\pi i T}$ is a very
convenient and systematic method for the construction of the
general threshold functions for any number of Wilson lines \cite{msp}.
The coefficient functions of powers of $q_T$ transform as
$n$--th order theta--functions in the various limits of only one
non--vanishing Wilson line modulus. The space of these functions is
known to be generated by a basic
set of theta functions \cite{mum1}. Modular invariance
in $U$ reduces this set considerably and the precise linear combination
of the few remaining generators is fixed by the additional symmetries.
For example the ${\cal O}(e^{-2\pi T_2})$ correction to \req{tinfty} can be
easily determined by this method to be
\be
-\h q_T \sum_{\al,\bet} \frac{\theta_\al^2\lf[\fc{B-C}{2},U\ri]\
\theta_\bet^2\lf[\fc{i(B+C)}{2},U\ri]}
{\theta_\al^2(0,U)\ \theta_\bet^2(0,U)} \ ,
\ee
where the $\theta_\al$ are the even $g=1$ theta functions and the sum is over
the six pairs $\al,\bet \in \{0,2,3\}, \ \al\neq \bet$.

To illustrate the formal derivation of the one--loop gauge couplings above,
let us give two concrete examples for our orbifold model. To fix the
coefficients we have calculated the $T\ra i\infty$ and $B\ra 0$ limits
directly from the integral representation \req{th8}. The agreement of the
functional dependence of these lowest order terms provides also a consistency
check of our formal derivation. An example for the case $I$ is the $SU(2)$
factor in the first $E_8$, broken to a $U(1)$ by the Wilson line. Any factor
from the $E_8^\prime$ falls into case $II$. For the corresponding one--loop
couplings we find:
\begin{eqnarray}
\triangle_{U(1)}^{I}&=&\fc{b_{U(1),0}^{N=2}}{10} \ln Y^{10} |\Cc_{10}|^2\ , \\
\triangle_{SU(4)'}^{II}&=&\fc{b_{SU(4)',0}^{N=2}}{12}\ln Y^{12}
|\Cc_{12}|^2\ .
\end{eqnarray}
\section{Conclusions and outlook}
The functional dependence of the one--loop gauge couplings
considered in this paper has been determined in a specific orbifold
compactification. However the previous experience with the dependence
of one--loop couplings on K\"ahler and complex structure moduli indicates
that these automorphic functions  should have a broader application.
This fact is a consequence of the general structure of duality symmetries
which causes a large stability of the functional dependence on the
moduli against variations of the specific  compactification \cite{ms1}
and even
the kind of interactions such as gauge, gravitational
and Yukawa couplings \cite{antl}.

An important question is that of the meaning of the Riemann surface
which gives rise to the automorphic functions representing the
one--loop gauge couplings. If there is any, one should find
analogous structures in the results for one--loop couplings in
general Calabi--Yau compactifications \cite{ber1},
generalizations to any number of Wilson lines and an explicit
representation of the mass formula for the exceptional states at
the enhanced symmetry points in terms of vanishing integrals of
certain abelian differentials. An interesting feature is the
resolution of the direct product structure\footnote{
A similar although different direct product structure plays a central
r\^ole in the derivation of the perturbative monodromies in
N=2 supersymmetric string theories in \cite{antn}.}
of the $T$ and $U$ dependent tori into a common Riemann surface
for non--vanishing vev of the Wilson line. Although increasing the
genus of the Riemann surface is certainly the wrong concept to
introduce additional Wilson lines, there is amazingly another candidate
ready to take the r\^ole of the Teichm\"uller parameter of an additional
handle: the dilaton. In fact eqs. \req{kp001}, \req{kp002} have an obvious
genus three generalization including the dilaton and the associated
modular forms of degree three represent automorphic functions
of the target space and S duality symmetry groups. An identical
situation is given in certain orbifold compactifications with
the dilaton replaced by a K\"ahler modulus.

As we have mentioned at the beginning the result for the moduli--dependent
threshold corrections of the N=1 supersymmetric orbifold compactification
immediately takes
over to the associated N=2 theory which consists of the N=2 sectors only.
The knowledge of the perturbative gauge coupling and the rich structure
of the moduli space make this N=2 theory an interesting subject to study
the generalization of strong--weak--coupling duality in N=2 globally
supersymmetric theories to local supersymmetry \cite{msp}.

Finally, moduli dependent threshold corrections to gauge couplings
in (0,2) orbifold models play an important r\^ole for the
discussion of gauge coupling unification. In contrary to moduli
independent threshold corrections \cite{kap1,ms3} they
can give rise to significant contributions.
A reasonable choice of the vevs of the moduli fields $T,U,B$ and $C$ should
allow for gauge coupling unification at $M_{\rm X}=2 \cdot 10^{16}{\rm GeV}$
without a grand unifying group \cite{mns2}.
\ \\ \\
{\bf Acknowledgments.}
We would like to thank Albrecht Klemm, Wolfgang Lerche,
Jan Louis, Hans Peter Nilles and  Stefan Theisen
for helpful discussions.\\ \\

\ \\

\newpage
\begin{figure}[h]
\vskip 2cm
\hskip 2cm
\psfig{figure=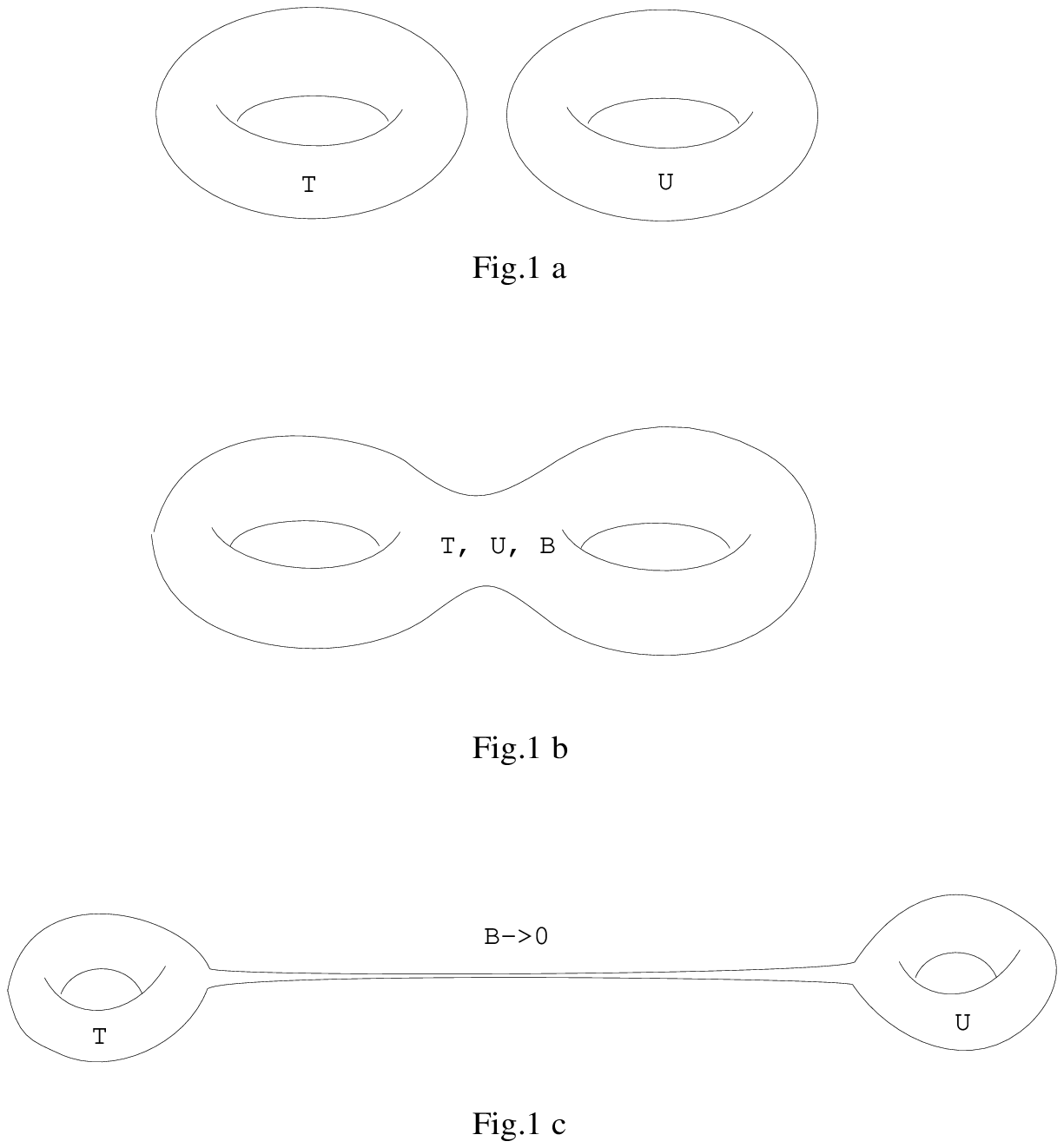}
\end{figure}

\ \\ \\ \\
{\bf Figure 1}: Riemann surfaces associated to the automorphic functions
representing the one--loop gauge couplings: {\em a}): direct product structure
corresponding to eq. \req{oldth}.
{\em b}): g=2 surface associated to the Wilson line
dependent coupling functions \req{del1},\req{del2} and \req{del32}.
{\em c}): degeneration limit of vanishing Wilson lines
replacing {\em a)}.
\end{document}